\documentclass[lettersize,journal]{IEEEtran}
\usepackage{amsmath,amsfonts}
\usepackage{algorithmic}
\usepackage{algorithm}
\usepackage{array}
\usepackage[caption=false,font=normalsize,labelfont=sf,textfont=sf]{subfig}
\usepackage{textcomp}
\usepackage{stfloats}
\usepackage{url}
\usepackage{verbatim}
\usepackage{graphicx}
\usepackage{cite}
\usepackage{bm} %% additional package
\usepackage{eucal}  %% additional package
\usepackage{boondox-cal} %% additional package
\usepackage{threeparttable} % additional package
\usepackage{multirow}
\usepackage{booktabs}  
\begin{document}

\title{SPHR-SAR-Net: Superpixel High-resolution SAR Imaging Network Based on Nonlocal Total Variation}
\author{Guoru Zhou,
	Zhongqiu Xu,
	Yizhe Fan,
	Zhe Zhang,~\IEEEmembership{Member,~IEEE,}
	Xiaolan Qiu,~\IEEEmembership{Senior Member,~IEEE,}
	Bingchen Zhang,
	Kun Fu
	and Yirong Wu
	
	\thanks{Corresponding author: Kun Fu, Email: fukun@aircas.ac.cn.}
	\thanks{G. Zhou, Z. Xu, Y. Fan, X. Qiu, B. Zhang, K. Fu and Y. Wu are with the Key Laboratory of Technology in Geo-spatial Information Processing and Application System, Chinese Academy of Sciences, Beijing, China.}
	\thanks{Z. Zhang and X. Qiu are also with Suzhou Key Laboratory of Microwave Imaging, Processing and Application Technology, and Suzhou Aerospace Information Research Institute, Suzhou, Jiangsu, China}
	\thanks{G. Zhou, Z. Xu, Y. Fan, Z. Zhang, X. Qiu, B. Zhang, K. Fu and Y. Wu are also with Aerospace Information Research Institute, Chinese Academy of Sciences, and School of Electronic, Electrical and Communication Engineering, University of Chinese Academy of Sciences, Beijing, China.}
	\thanks{Manuscript received Nov xx, 20xx; revised August xx, 20xx.}
}

\markboth{IEEE JOURNAL OF SELECTED TOPICS IN APPLIED EARTH OBSERVATIONS AND REMOTE SENSING,~Vol.~14, No.~8, February~2023}%
{Shell \MakeLowercase{\textit{et al.}}: A Sample Article Using IEEEtran.cls for IEEE Journals}

\IEEEpubid{00000000000000000000000000000000000000000000}

\maketitle

\begin{abstract}
High-resolution is a key trend in the development of synthetic aperture radar (SAR), which enables the capture of fine details and accurate representation of backscattering properties. However, traditional high-resolution SAR imaging algorithms face several challenges. Firstly, these algorithms tend to focus on local information, neglecting non-local information between different pixel patches. Secondly, speckle is more pronounced and difficult to filter out in high-resolution SAR images. Thirdly, the process of high-resolution SAR imaging generally involves high time and computational complexity, making real-time imaging difficult to achieve. To address these issues, we propose a Superpixel High-Resolution SAR Imaging Network (SPHR-SAR-Net) for rapid despeckling in high-resolution SAR mode.
Based on the concept of superpixel techniques, we initially combine non-convex and non-local total variation as compound regularization. This approach more effectively despeckles and manages the relationship between pixels while reducing bias effects caused by convex constraints. Subsequently, we solve the compound regularization model using the Alternating Direction Method of Multipliers (ADMM) algorithm and unfold it into a Deep Unfolded Network (DUN). The network's parameters are adaptively learned in a data-driven manner, and the learned network significantly increases imaging speed. Additionally, the Deep Unfolded Network is compatible with high-resolution imaging modes such as spotlight, staring spotlight, and sliding spotlight.
In this paper, we demonstrate the superiority of SPHR-SAR-Net through experiments in both simulated and real SAR scenarios. The results indicate that SPHR-SAR-Net can rapidly perform high-resolution SAR imaging from raw echo data, producing accurate imaging results.
\end{abstract}


\section{Introduction}
\IEEEPARstart{S}{ynthetic} aperture radar (SAR) is an active imaging system. Unlike optical sensors, SAR transmits microwave signals with surface penetration capabilities, enabling it to function in all weather and day-night conditions. Consequently, SAR is widely used in geographical mapping, resource exploration, target surveillance, and other fields \cite{ccetin2014sparsity}. High resolution is an essential trend in the development of advanced SAR systems.

Compared with low-resolution SAR images, high-resolution SAR images offer greater detail and enhance the accuracy of various measurements, thereby improving target identification and recognition \cite{ouchi2013recent}. To increase the resolution of SAR images, several SAR operation modes have been proposed, such as spotlight SAR and sliding spotlight SAR. The spotlight and sliding spotlight modes provide longer aperture times, resulting in more data being collected \cite{lanari2001new}. However, traditional matched-filter-based imaging algorithms for these operating modes suffer from speckle, sidelobes, and noise, which affect the quality of SAR images \cite{raney1994precision}. Additionally, complex matrix multiplication makes the imaging process intricate, with high time and computational complexity. Therefore, devising accurate and efficient high-resolution SAR imaging methods is a significant research topic.

With the improvement of SAR image resolution, we can now capture the fine details of target objects. In high-resolution SAR images, most target objects consist of multiple physical pixels, and there is a substantial connection and correlation among neighboring pixels belonging to the same target. We refer to this phenomenon as "superpixel," indicating that it contains considerably more information than conventional physical pixels. In the SAR imaging context, the superpixel concept is realized by breaking down each imaging pixel into smaller subpixels, each representing a different value. By exploiting the values of adjacent pixels, it is possible to interpolate the value of pixels at specific locations using correlation information among neighboring pixels, ultimately creating SAR images with finer details and richer information.

To implement the superpixel concept, there are three challenges that need to be addressed. First, the speckle of SAR images becomes more pronounced. In high-resolution SAR images, the pixel size is smaller, meaning that each pixel contains fewer scatterers. Speckle is caused by interference between scatterers, and it can be more pronounced in smaller pixels. Second, due to the increased number of pixels within a unit, local information becomes more salient, while nonlocal information is often ignored. Nonlocal techniques consider the similarity of patches within the image, assuming that patches similar to one another are likely to be part of the same object \cite{gilboa2009nonlocal,liu2014new}. Third, as pixel size decreases, the amount of data 
\\ \hspace*{\fill} \\
increases, making the processing of raw data more complex. As a result, the imaging process faces high computational complexity and time complexity.

In recent years, high-resolution SAR imaging methods based on compressed sensing have gradually matured \cite{ender2010compressive,zhang2012sparse,baraniuk2007compressive,dong2013novel}. In \cite{ccetin2001feature}, the authors employ regularization techniques to incorporate prior information about interesting features into the imaging formation. Subsequently, numerous studies have enhanced SAR imaging results with convincing outcomes by adding various regularization constraints to highlight different target characteristics \cite{wei2019improved,bi2016lq,zhao2014adaptive,zhang2013compressive,xu2022sparse}.

To better address the challenges in high-resolution SAR imaging, non-local total variation (NLTV) has gained attention \cite{werlberger2010motion}. The NLTV model combines the advantages of the total variation (TV) model \cite{chambolle2004algorithm} and the non-local mean (NLM) method \cite{protter2008generalizing}. It can more effectively process image texture and edge information while considering non-local information to achieve better image quality \cite{zhang2013compressive}. Additionally, to reduce the bias effect due to convex constraints and improve the accuracy of high-resolution SAR imaging, non-convex constraints are introduced as compound regularization terms \cite{xu2022nonconvex}.

However, iterative-based regularization algorithms generally involve a large number of iterations and extensive computation, along with manual parameter setting. As a result, an increasing number of data-driven SAR imaging methods have been proposed. Compared to convolutional neural network (CNN)-based methods \cite{mason2017deep,gao2018enhanced,pu2021deep}, deep unfolding networks (DUN) are designed around the iterative updating process of the applied algorithm, and learnable parameters in the network are automatically obtained through training, providing better interpretability \cite{monga2021algorithm}. ADMM-CSNet was first applied in magnetic resonance imaging (MRI) image reconstruction, demonstrating the superiority of DUN \cite{yang2018admm}. Several works have since used DUN for SAR imaging. Based on the iterative shrinkage threshold algorithm (ISTA), the authors proposed a deep unfolding network with approximated observation to learn the imaging procedure from radar echoes \cite{kang2022sar,zhao2021end,zhang2022sr}. Additionally, the alternating direction method of multipliers (ADMM) framework is more flexible and suitable for diverse regularization. In \cite{wei2021sar}, PSRI-Net was proven to enhance point scatterers, line-segment scatterers, and rectangular plate scatterers. In \cite{li2022target}, MF-ADMM-Net was used to enhance targets from background clutters. STLS-LADMM-Net \cite{li2022stls} can perform phase compensation during imaging for autofocus imaging. LRSR-ADMM-Net \cite{an2022lrsr} combined low-rank and sparse regularization as a joint constraint for imaging non-sparse observation scenes with reduced data. Furthermore, SPB-Net can simultaneously perform imaging and denoising \cite{xiong2020spb,xiong2021sar,xiong2021q}. In \cite{zhang2021learning}, the authors despeckle SAR images using sparse representation and deep learning techniques. However, these methods do not specifically address the characteristics of high-resolution SAR images and lack pertinence to high-resolution SAR imaging.

In this paper, we propose a superpixel high-resolution SAR network for fast imaging and despeckling. Drawing on regularization techniques, we employ nonlocal total variation to impose global constraints on pixels for despeckling and incorporate nonconvex regularization as a joint constraint. Subsequently, we use the deep unfolding alternating direction method of multipliers network to solve the aforementioned joint compound regularization. The network parameters are trained in a data-driven manner, mitigating the impact of empirical parameter setting on image quality. Furthermore, our proposed method is suitable for various high-resolution SAR imaging modes. We demonstrate the superiority of the proposed method through simulation experiments and real data experiments. The main contributions of our work can be summarized as follows:

\begin{itemize}
	\item We thoroughly analyze the challenges of the high-resolution SAR imaging process. To address these issues, we introduce superpixel processing, despeckling through NLTV constraints, and restoring radiometric accuracy using non-convex constraints.
	\item We employ the ADMM unfolding network method to solve the aforementioned model, resulting in fast and accurate high-resolution SAR imaging. The network parameters are trained in a data-driven manner, overcoming the difficulties of parameter adaptation, high computational complexity, and high time complexity associated with conventional iterative methods.
	\item The proposed method is suitable for multiple high-resolution imaging modes, such as spotlight, staring spotlight, and sliding spotlight, making the approach more widely applicable.
\end{itemize}

The remainder of this article is organized as follows. Section \ref{model} offers a detailed analysis of the challenges encountered in high-resolution SAR imaging and establishes a superpixel-based imaging model for high-resolution imaging modes. Section \ref{network} presents the solution to the imaging model and elaborates on the details of the superpixel high-resolution SAR imaging network (SPHR-SAR-Net). Experimental results and discussion are provided in Section \ref{experiment}. Conclusions and contributions are presented in Section \ref{conclusion}.

\begin{figure}[!t]  %% [!t]在该页的顶部
	\centering
	\includegraphics[width=2.5in]{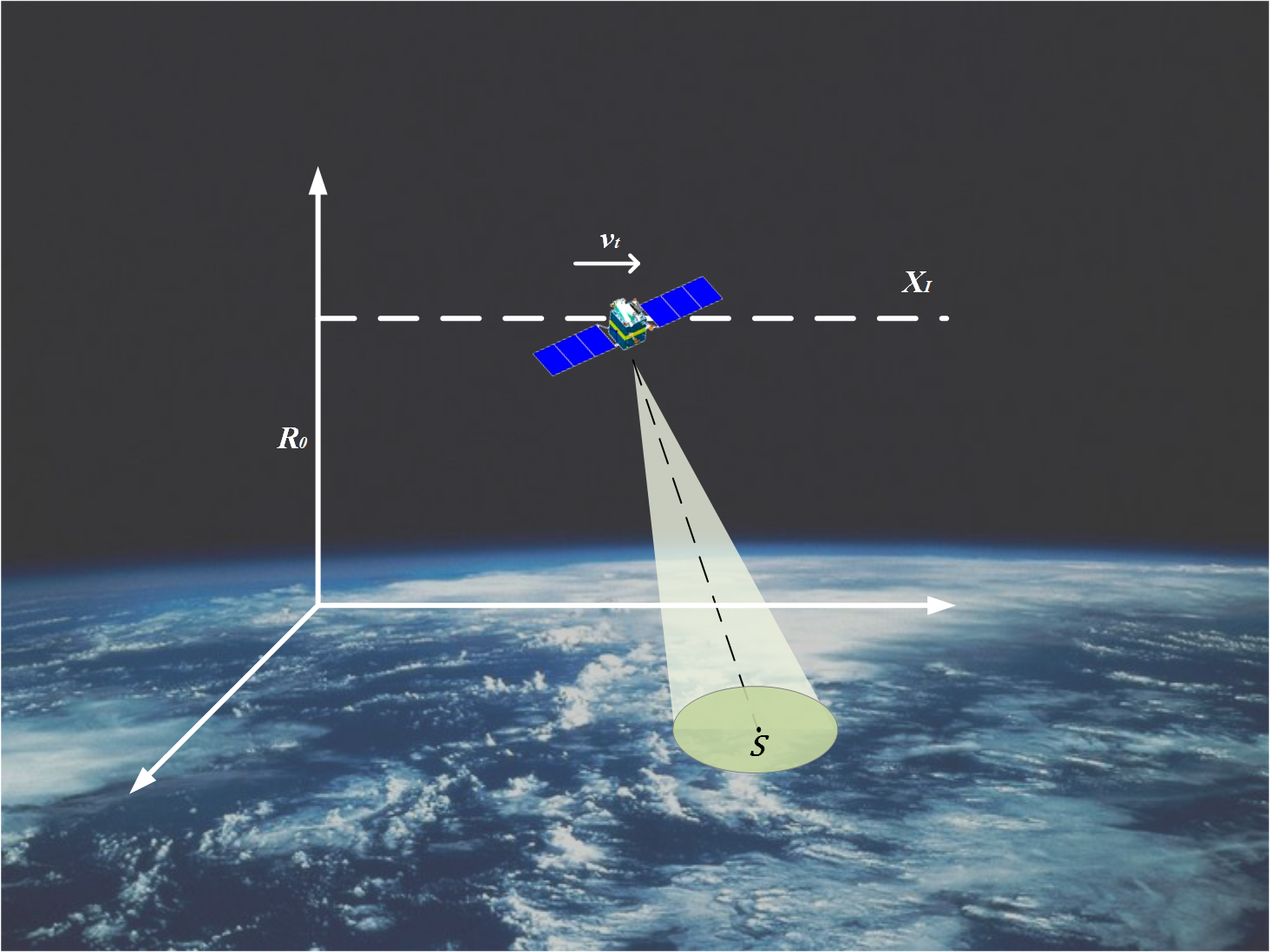}
	\caption{The high-resolution spaceborne SAR imaging geometry model of spotlight SAR.}
	\label{fig1}
\end{figure}

\section{High-Resolution SAR Imaging Model} \label{model}
\subsection{High-Resolution Sparse SAR Imaging Model}
According to the theory of sparse microwave imaging \cite{ccetin2014sparsity,zhang2012sparse} the SAR observation model can be expressed as a linear model:
\begin{equation}
	\emph{\textbf{y}} = \bm{\Phi} \emph{\textbf{x}}+\emph{\textbf{n}}
\end{equation}
where $\emph{\textbf{x}} = vec\left(\textbf{X}\right) \in \mathbb {C}^{N \times 1}$, $\textbf{X} \in \mathbb {C}^{N_{p} \times N_{q}}$ is the backscattering coefficient matrix of the two-dimensional surveillance scene, $N={N_{p} \times N_{q}}$. $\emph{\textbf{y}} = vec\left(\textbf{Y}\right) \in \mathbb {C}^{M \times 1}$, $\textbf{Y} \in \mathbb {C}^{N_{a} \times N_{r}}$ is the two-dimensional echo matrix, $M={N_{a} \times N_{r}}$. $\bm{\Phi} \in \mathbb {C}^{M \times N}$ is the observation matrix of a high-resolution SAR system. Commonly used high-resolution imaging modes for spaceborne SAR include spotlight, staring spotlight, and sliding spotlight. The resolution of advanced spotlight mode SAR satellites reaches up to 1 meter, and the resolution of advanced staring spotlight mode SAR satellites goes down to 0.25 meters \cite{TerraSAR-X}. The high-resolution spaceborne SAR imaging geometry model for spotlight SAR is depicted in Fig. \ref{fig1}.

The observation matrix can be described by the imaging geometric relationship. By discretizing the time domain $t_{m}$ and the geometric space domain $x_n$, the observation matrix can be expressed as:
\begin{equation}
	\begin{aligned}
	\bm{\Phi} \left( m,n \right) &= rect \left( \frac{{{v}_{t}}{{t}_{m}}}{{{X}_{I}}} \right)rect\left( \frac{A{{v}_{t}}{{t}_{m}}-{{x}_{n}}}{X}\right) \\ 
     & \times rect\left( \frac{t-{2R\left( {{t}_{m}} \right)}/{c}\;}{{{T}_{p}}} \right) \exp\left( -j\frac{4\pi R\left( {{t}_{m}} \right)}{\lambda } \right)\\
     & \times \exp\left( j\pi {{K}_{r}}{{\left( t-\frac{2R\left( {{t}_{m}} \right)}{c} \right)}^{2}} \right) \\ 
	\end{aligned}
\end{equation}
where ${v}_{t}$ is the velocity of the SAR system, ${X}_{I}$ is the length of the platform track, $X$ is the length of the imaging scene, ${T}_{p}$ and ${\lambda }$  is the pulse width and the wavelength of the transmitted linear frequency modulation (LFM) signal $R\left( {{t}_{m}} \right)$ is the instantaneous slope distance at ${t}_{m}$, $A$ is the sliding factor, in which $A=0$, $A=1$ and $0<A<1$ are the spotlight SAR, stripmap SAR and sliding spotlight SAR, respectively.

The high-resolution SAR imaging procedure is to get unknown $\emph{\textbf{x}}$ through known $\emph{\textbf{y}}$ and $\bm{\Phi}$, which can be regarded as a linear inverse problem (LIP). We can derive $\hat{\emph{\textbf{x}}}$ by minimizing the regularized linear least-squares cost function:
\begin{equation}
	\hat{\emph{\textbf{x}}}=\mathop{\arg}\mathop{\min}\limits_{\emph{\textbf{x}}} \left\| \emph{\textbf{y}}-\bm{\Phi} \emph{\textbf{x}} \right\|_{2}^{2}+\sum\limits_{n}{{{\lambda }_{n}}{{p}_{n}}\left( \emph{\textbf{x}} \right)}
\end{equation}
where ${\lambda }_{n}$ is the $nth$ regularization parameter, ${{p}_{n}}\left( \emph{\textbf{x}} \right)$ is the $n$-th regularization function. For different prior features, they can be enhanced by different regularizations. In conventional SAR sparse microwave imaging, $\ell_1$ norm regularization is used to constrain the sparsity of the image, and TV norm regularization is used to constrain the sparsity of the gradient field for despeckling. However, in high-resolution SAR images, the structural information of scatterers becomes more refined, necessitating the application of new regularization techniques to more accurately constrain image features.

\subsection{Challenges in High-Resolution SAR Imaging}
\begin{figure}[!t]
	\centering
	\subfloat[]{\includegraphics[width=1in]
		{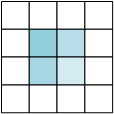} \label{low_resolution}}
	\hfil
	\subfloat[]{\includegraphics[width=1in]
		{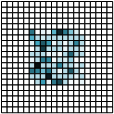} \label{high_speckle}}
	\hfil
	\subfloat[]{\includegraphics[width=1in]
		{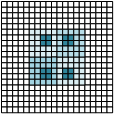} \label{high_resolution}}
	\caption{SAR image schematic diagram in different resolution. (a) Low-resolution SAR image; (b) High-resolution SAR image with speckle; (c) High-resolution SAR image without speckle.}
	\label{fig2}
\end{figure}

High resolution is a crucial trend in the development of Synthetic Aperture Radar (SAR) systems \cite{ouchi2013recent}. High-resolution SAR images offer increased detail and accuracy, enabling more precise and comprehensive analysis of target scenes. Identifying and analyzing features of interest within the target scene becomes significantly easier with high-resolution images. Additionally, these images facilitate the enhanced discrimination of features, such as differences in surface roughness or vegetation cover. Furthermore, high-resolution SAR images can detect smaller targets that might otherwise merge into a single blob in low-resolution images, as illustrated in Fig. \ref{low_resolution}. This capability is particularly important in applications such as maritime surveillance.

Despite the numerous advantages of high-resolution images, several challenges arise in high-resolution SAR imaging, including:
\subsubsection{Speckle}
Speckle noise arises from the coherent nature of the radar signal and the random interference between waves reflected from the target, which can obscure or distort target information in the image \cite{xiong2020spb}. However, speckle noise is more prominent in high-resolution SAR images, as shown in Fig. \ref{high_speckle}. Due to the smaller pixel size and increased radiometric accuracy, high-resolution SAR images are more sensitive to subtle changes in the target scene. This results in an increased appearance of speckle noise in the image, particularly in areas with high texture.
 
\subsubsection{Image Interpretation}
As the pixel size decreases, targets are represented by more pixels, causing people to focus more on local information while overlooking nonlocal information in the image. Nonlocal information provides crucial context for the target area, enabling more accurate interpretation of the target scene. Moreover, nonlocal information can be used to enhance the quality of local information in high-resolution SAR images. Techniques like despeckling depend on surrounding regions to offer context for removing speckle from the SAR image, as illustrated in Fig. \ref{high_resolution} \cite{protter2008generalizing}. 
   
\subsubsection{Imaging Algorithm}
The raw data amount of high-resolution SAR is huge, making it challenging to process. Handling high-resolution SAR images necessitates sophisticated algorithms capable of extracting valuable information from complex data. These algorithms are often computationally demanding, and parameter setting can be intricate.

\subsection{NLTV and nonconvex comnpound regularization}
Based on the above analysis, nonlocal total variation constraints are applied to high-resolution SAR imaging \cite{xu2020accurate}. Assuming the magnitude of the high-resolution SAR image is expressed as $\left| \textbf{X} \right|\in {{\mathbb{R}}^{{{N}{p}}\times {{N}{q}}}}$, the nonlocal gradient at pixel $i$ can be characterized by \cite{liu2014new}:
\begin{equation}
	{{\nabla }_{NLTV}}{{\left| \textbf{X} \right|}_{i}}=\sqrt{w\left( i,j \right)}\left( {{\left| \textbf{X} \right|}_{i}}-{{\left| \textbf{X} \right|}_{j}} \right) \ i,j\in \left| \textbf{X} \right|
\end{equation}
where $w\left( i,j \right)$ is the weighting function to describe the coherence between two patches centered at pixel $i$ and pixel $j$ . In this paper, $w\left( i,j \right)$ is described as the Gaussian weighting function:
\begin{equation}
	w\left( i,j \right)= \exp \left\{ -\frac{\sum\limits_{d}{{{\operatorname{\textbf{G}}}_{\sigma }}\left( d \right)\left| \left| {{\left| {\mathbf{\hat{\textbf{X}}}} \right|}_{i,d}}-{{\left| {\mathbf{\hat{\textbf{X}}}} \right|}_{j,d}} \right| \right|_{2}^{2}}}{2{{h}^{2}}} \right\}
\end{equation}
where $\textbf{G}_{\sigma }\left( d \right) $  is the Gaussian Kernel with standard deviation in the patch size $d$, $h$ is the filtering parameter to control the smoothness. The NLTV regularization of high-resolution SAR image is formulated as:
\begin{equation}
	{{p}_{NLTV}}\left(  \emph{\textbf{x}} \right)={{\left\| |{{\nabla }_{NLTV}}\left| \textbf{X} \right|| \right\|}_{1}}
\end{equation}

To leverage the sparse prior of high-resolution SAR images and minimize the bias effect to achieve accurate radiometric accuracy, we simultaneously incorporate a nonconvex constraint as regularization. The generalized minimax-concave (GMC) penalty, a commonly used nonconvex penalty, can be expressed as \cite{selesnick2017sparse}:
\begin{equation}
	{{p}_{GMC}}\left(  \emph{\textbf{x}} \right)=\left\|  \emph{\textbf{x}} \right\|-\underset{ \emph{\textbf{v}} \in \mathbb{R}}{\mathop{\min }}\,\{\left\|  \emph{\textbf{v}} \right\|-\frac{1}{2}\left\| \mathbf{B}( \emph{\textbf{x}}- \emph{\textbf{v}}) \right\|_{2}^{2}\}
\end{equation}
where $\mathbf{B}$ is the weighting matrix.

In summary, the high-resolution SAR imaging problem can be modeled as a compound regularization problem with NLTV and nonconvex constraints, which is given by:
\begin{gather}
		\mathop{\arg}\mathop{\min}\limits_{\emph{\textbf{x}}} \ {{\lambda }_{NLTV}}{{p}_{NLTV}}\left( \emph{\textbf{x}} \right) +  {{\lambda }_{NC}}{{p}_{NC}}\left( \emph{\textbf{x}} \right)   \nonumber\\ 
		 \mathrm{s.t.} \ \emph{\textbf{y}} = \bm{\Phi} \emph{\textbf{x}}+\emph{\textbf{n}}
		 \label{imaging_model}
\end{gather}
Equ. (\ref{imaging_model}) can be written in the form of unconstrained optimization as:
\begin{equation}
	\begin{aligned}
	\hat{\emph{\textbf{x}}}&=\mathop{\arg}\mathop{\min}\limits_{\emph{\textbf{x}}} \left\| \emph{\textbf{y}}-\bm{\Phi} \emph{\textbf{x}} \right\|_{2}^{2}  \\ &+{{{\lambda }_{NLTV}}{{p}_{NLTV}}\left( \emph{\textbf{x}} \right)}+{{{\lambda }_{NC}}{{p}_{NC}}\left( \emph{\textbf{x}} \right)} \\
	\end{aligned}
\end{equation}
	
\subsection{High-Resolution SAR Fast Imaging Model}

\begin{figure}[!t]
	\centering
	\subfloat[]{\includegraphics[width=2.5in]
		{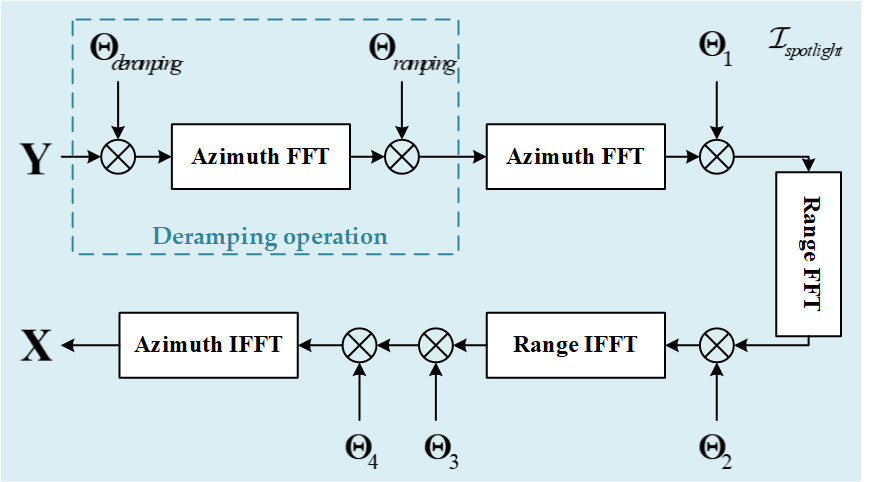} \label{Imaging_operater}}
	\hfil
	\subfloat[]{\includegraphics[width=2.5in]
		{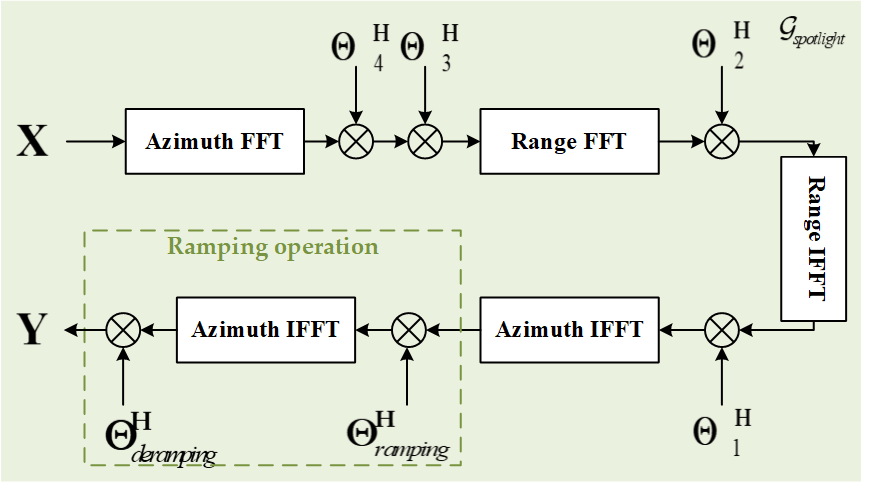} \label{Inverse_Imaging_operater}}
	\hfil
	\caption{High-resolution SAR Fast Imaging Operators. (a) Spotlight/Sliding spotlight SAR imaging operator; (b) Spotlight/Sliding spotlight SAR inverse imaging operator.}
	\label{fig3}
\end{figure}

\begin{figure*}[!t]
	\begin{equation}
		{\mathcal{I}_{spotlight}}\left( \textbf{Y} \right)=\textbf{F}_{a}^{-1}\left(\textbf{F}_{r}^{-1}\left( {\textbf{F}_{r}}\left( {\textbf{F}_{a}}\left( {\textbf{F}_{a}}\left( \textbf{Y} \odot {{ \bm{\Theta} }_{deramping}} \right)\odot {{\bm{\Theta} }_{ramping}} \right) \odot {{\bm{\Theta} }_{1}} \right)\odot {{\bm{\Theta} }_{2}} \right)\odot {{\bm{\Theta} }_{3}}\odot {{\bm{\Theta}}_{4}} \right)
		\label{imaging_operater}
	\end{equation}
	
	\begin{equation}
		{\mathcal{G}_{spotlight}}\left( \textbf{X} \right)=\textbf{F}_{a}^{-1}\left( \textbf{F}_{a}^{-1}\left( \textbf{F}{r}^{-1}\left( {{\textbf{F}}_{r}}\left( {{\textbf{F}}_{a}}\left( \textbf{x} \right)\odot \bm{\Theta} _{4}^{H}\odot \bm{\Theta} _{3}^{H} \right)\odot \bm{\Theta} _{2}^{H} \right)\odot \bm{\Theta} _{1}^{H} \right)\odot \bm{\Theta} _{ramping}^{H} \right)\odot \bm{\Theta} _{deramping}^{H}
		\label{inverse_imaging_operater}
	\end{equation}
	\hrule 
\end{figure*}

Traditional imaging methods based on matched filtering exhibit severe sidelobe interference, particularly in high-resolution SAR images. In the case of compressed sensing methods, the construction of the observation matrix necessitates explicit imaging geometry. The substantial storage demands and computational complexity present practical challenges for raw data processing. Consequently, it is essential to employ multimode approximate measurement operators for rapid imaging \cite{fang2013fast}. In \cite{li2022target,xiong2020spb}, the authors have applied matched filter-based fast imaging operators with iterative solutions and deep unfolding networks in stripmap SAR. In this paper, we consider additional high-resolution SAR imaging modes.

For spotlight SAR and sliding spotlight SAR, the two-step algorithm \cite{lanari2001new} incorporates a deramping operation in matched filtering. As a result, the approximate measurement procedure for spotlight/sliding spotlight SAR and its inverse procedure are depicted in Fig. \ref{Imaging_operater} and Fig. \ref{Inverse_Imaging_operater}, respectively.

The spotlight/sliding spotlight SAR imaging operator and the inverse imaging operator can be expressed as Equ. (\ref{imaging_operater}) and Equ. (\ref{inverse_imaging_operater}), respectively, as described in \cite{zhou2022azimuth}. Here, $\textbf{F}_{a}$ and $\textbf{F}_{a}^{-1}$, as well as $\textbf{F}_{r}$ and $\textbf{F}_{r}^{-1}$, correspond to the Fourier and inverse Fourier transform matrices in azimuth and range, respectively; $\bm{\Theta} _{1}$ represents the differential RCMC matrix; $\bm{\Theta} _{2}$ is the bulk RCMC and matched filter matrix in range; $\bm{\Theta} _{3}$ denotes the matched filter matrix in azimuth; $\bm{\Theta} _{4}$ is the residual Doppler phase correction matrix; $\bm{\Theta} _{deramping}$ and $\bm{\Theta} _{ramping}$ are the deramping and ramping operation matrices, respectively; $\odot$ denotes the Hadamard product; and $\left( \cdot \right)^{H}$ represents the conjugate transpose operator.

By employing the approximate measurement operators and taking into account the azimuth and range downsampling, the high-resolution SAR observation model can be expressed in a two-dimensional form:
\begin{equation}
	\begin{aligned}
		\hat{{\textbf{X}}}&=\mathop{\arg}\mathop{\min}\limits_{{\textbf{X}}} \left\| {\textbf{Y}}-\bm{\Theta}_{a} {\mathcal{G}\left( \textbf{X} \right) } \bm{\Theta}_{r} \right\|_{2}^{2}  \\ &+{{{\lambda }_{NLTV}}{{p}_{NLTV}}\left({\textbf{X}} \right)}+{{{\lambda }_{NC}}{{p}_{NC}}\left({\textbf{X}} \right)} \\
	\end{aligned}
    \label{solving_model}
\end{equation}
where $\bm{\Theta}_{a}$ and $\bm{\Theta}_{r}$ are the azimuth and range down-sampling matrices, respectively.

\section{SPHR-SAR-Net for \\ High-Resolution SAR Imaging} \label{network}
\subsection{Iterative Solution of High Resolution SAR Imaging Model} \label{iterative_solution}
To solve compound regularization optimization problems, the ADMM algorithm demonstrates exceptional performance \cite{guven2016augmented}. In comparison with other solution methods, ADMM is versatile and adaptable for various resolutions. Inspired by the modified variable splitting ADMM, the modified constrained form of Equ. (\ref{solving_model}) can be represented as:
\begin{multline*}
	\mathop{\min}\limits_{{\textbf{X}}}\left\| \textbf{Y}-{{\bm{\Theta}}_{a}}\mathcal{G}\left( \textbf{X} \right){\bm{\Theta }_{r}} \right\|_{2}^{2} \\
	+{{\lambda }_{NLTV}}{{p}_{NLTV}}\left( {\textbf{Z}_{1}} \right)+{{\lambda }_{GMC}}{{p}_{GMC}}\left( {\textbf{Z}_{2}} \right) \nonumber 
\end{multline*}
\begin{gather}
		\mathrm{s.t.} \ \textbf{G}\textbf{X}=\textbf{Z},\textbf{G}={{\left[ {{\textbf{I}}^{T}},{{\textbf{I}}^{T}} \right]}^{T}},\textbf{Z}={{\left[ {\textbf{Z}_{1}}^{T},{\textbf{Z}_{2}}^{T} \right]}^{T}} 
\end{gather}
where $\textbf{Z}_{1}$ and $\textbf{Z}_{2}$ are auxiliary variables, $\textbf{I}$ is the identity matrix. The augmented Lagrangian function of the above constrained optimization problem is Equ. (\ref{Lagrangian_function})
\begin{figure*}[hb]
	\hrule 
  \begin{equation}
	\begin{aligned}
		{{\mathcal{L}}_{\rho }}\left( \textbf{X},\textbf{Z},\bm{\mu}  \right)
		& =\left\| \textbf{Y}-{\bm{\Theta }_{a}}\mathcal{G}\left( \textbf{X} \right){\bm{\Theta }_{r}} \right\|_{2}^{2}+{{\lambda }_{NLTV}}{{p}_{NLTV}}\left( {\textbf{Z}_{1}} \right)+{{\lambda }_{GMC}}{{p}_{GMC}}\left( {\textbf{Z}_{2}} \right) +\left\langle \bm{\mu} ,\textbf{Z}-\textbf{GX} \right\rangle +\frac{\rho }{2}\left\| \textbf{GX}-\textbf{Z} \right\|_{2}^{2} \\
		& =\left\| \textbf{Y}-{\bm{\Theta }_{a}}\mathcal{G}\left( \textbf{X} \right){\bm{\Theta }_{r}} \right\|_{2}^{2}+{{\lambda }_{NLTV}}{{p}_{NLTV}}\left( {\textbf{Z}_{1}} \right)+{{\lambda }_{GMC}}{{p}_{GMC}}\left( {\textbf{Z}_{2}} \right)
		+\frac{\rho }{2}\left\| \textbf{GX}-\textbf{Z}-\textbf{D} \right\|_{2}^{2}+{C}  
	\end{aligned}
	\label{Lagrangian_function}
  \end{equation}
\end{figure*}
, where $\textbf{D}= \frac{1}{\rho} \bm{\mu} $ is the scaled augmented Lagrangian multipliers, as for compound regularization $\textbf{D}={{\left[ {\textbf{D}_{1}}^{T},{\textbf{D}_{2}}^{T} \right]}^{T}}   $, $\bm{\mu}$ is the augmented Lagrangian multipliers, $\rho>0$  is the penalty parameter, and ${C}$ is a constant independent of optimization variables.

According to ADMM algorithm, the iterative optimization process of Equ. (\ref{Lagrangian_function}) can be described in detail as:

\subsubsection{Reconstruction part $\emph{\textbf{X}}^{(t+1)}$ updating step}
\begin{multline}
	{\textbf{X}^{\left( t+1 \right)}} = \mathop{\arg}\mathop{\min}\limits_{{\textbf{X}}}\left\| \textbf{Y}-{\bm{\Theta }_{a}}\mathcal{G}\left( {\textbf{X}^{\left( t \right)}} \right){\bm{\Theta }_{r}} \right\|_{2}^{2} \\
	 +\frac{\rho }{2}\left\| \textbf{G}{{\textbf{X}}^{\left( t \right)}}-{{\textbf{Z}}^{\left( t \right)}}-{{\textbf{D}}^{\left( t \right)}} \right\|_{2}^{2}
	 \label{x_update}
 \end{multline}
Since the observation matrix of a high-resolution SAR system can be approximated as a unitary matrix, and the SAR imaging operator and inverse imaging operator can be considered approximately reversible \cite{li2022target}, Equ. (\ref{x_update}) can be reformulated as:
\begin{multline}
	{\textbf{X}^{\left( t+1 \right)}}= \\
	\mathcal{I}\left\{ \bm{\Gamma} \odot \left[ {\bm{\Xi }_{a}}\odot \textbf{Y}\odot {\bm{\Xi }_{r}}+\mathcal{G}\left\{ \rho {{\textbf{G}}^{T}}\left( {\textbf{Z}^{\left( t \right)}}-{\textbf{D}^{\left( t \right)}} \right) \right\} \right] \right\}
	\label{X_solution}
\end{multline}
where $\bm{\Xi }_{a}$ and $\bm{\Xi }_{r}$ is the random sampling matrix defined
by $\bm{\Theta }_{a}$ and $\bm{\Theta }_{r}$, $\bm{\Gamma}$ is the random downsampling indicator matrix determined by $\bm{\Xi }_{a}$ and $\bm{\Xi }_{r}$.

\subsubsection{NLTV part $\emph{\textbf{Z}}_{1}^{(t+1)}$ updating step} \label{NLTV_iteration}
\begin{multline}
	{\textbf{Z}_{1}}^{\left( t+1 \right)} = \mathop{\arg}\mathop{\min}\limits_{{\textbf{Z}_{1}}}{{\lambda }_{NLTV}}{{p}_{NLTV}}\left( {{\textbf{Z}_{1}}^{\left( t \right)}} \right)  \\
	+\frac{\rho }{2}\left\| {{\textbf{X}}^{\left( t+1 \right)}}-{{\textbf{Z}_{1}}^{\left( t \right)}}-{{\textbf{D}_{1}}^{\left( t \right)}} \right\|_{2}^{2}
	\label{NLTV_update}
\end{multline}
Equ. (\ref{NLTV_update}) can be solve by the modified Chambolle algorithm \cite{gilboa2009nonlocal}, and the solution form can be written as:
\begin{multline}
	{\textbf{Z}_{1}}^{\left( t+1 \right)}=\rm{sign} \left( {\textbf{X}^{\left( t+1 \right)}}-{\textbf{D}_{1}}^{\left( t \right)} \right)  \\
	\odot \left( \left| {\textbf{X}^{\left( t+1 \right)}}-{\textbf{D}_{1}}^{\left( t \right)} \right|-{{\lambda }_{NLTV}}{{\nabla }_{NLTV}}\left( {{\textbf{P}}^{\left( t+1 \right)}} \right) \right)
	\label{NLTV_solution}
\end{multline}
where ${{\nabla }_{NLTV}}(\cdot)$ is the nonlocal divergence operator, ${{\textbf{P}}^{\left( t+1 \right)}}$ is the auxiliary variable as:
\begin{multline}
	{\textbf{P}}^{\left( t+1 \right)} =  \\
	\frac{{\textbf{P}}^{\left( t \right)}+\tau {{\left( {{\Delta }_{NLTV}}\left( {{\nabla }_{NLTV}}\left( {\textbf{P}}^{\left( t \right)} \right)-\frac{\left| {{\textbf{X}}}^{\left( t+1 \right)}-{{\textbf{D}}}^{\left( t \right)} \right|}{{{\lambda }_{NLTV}}} \right) \right)}}}{1+\tau \left| {{\left( {{\Delta }_{NLTV}}\left( {{\nabla }_{NLTV}}\left( {\textbf{P}}^{\left( t \right)} \right)-\frac{\left| {{\textbf{X}}}^{\left( t+1 \right)}-{{\textbf{D}}}^{\left( t \right)} \right|}{{{\lambda }_{NLTV}}} \right) \right)}} \right|}
\end{multline}
where ${{\Delta }_{NLTV}}(\cdot)$ is the nonlocal gradient operator, $\tau$ is the nonlocal coefficient and $0<\tau<\frac{1}{\left\|  {{\nabla }_{NLTV}}(\textbf{P}^{\left( t \right) })\right\|_{2}^{2}} $.

\subsubsection{NC part $\emph{\textbf{Z}}_{2}^{(t+1)}$ updating step}
\begin{multline}
	{\textbf{Z}_{2}}^{\left( t+1 \right)} = \mathop{\arg}\mathop{\min}\limits_{{\textbf{Z}_{2}}}{{\lambda }_{NC}}{{p}_{NC}} \left( {{\textbf{Z}_{2}}^{\left( t \right)}} \right)  \\
	+\frac{\rho }{2}\left\| {{\textbf{X}}^{\left( t+1 \right)}}-{{\textbf{Z}_{2}}^{\left( t \right)}}-{{\textbf{D}_{2}}^{\left( t \right)}} \right\|_{2}^{2}
	\label{NC_update}
\end{multline}
Equ. (\ref{NC_update}) can be solve by the GMC threshold function \cite{wei2019improved}, and the solution form can be written as:
\begin{equation}
	{\textbf{Z}_{2}}^{\left( t+1 \right)} = {{\mathcal{S}}_{NC}}\left( {\textbf{X}^{\left( t+1 \right)}}-{\textbf{D}_{2}}^{\left( t \right)}, \ {{{\lambda }_{NC}}}/{\rho } \right)
	\label{NC_solution}
\end{equation}
where $\mathcal{S}$ is the GMC threshold function and the description is in Section \ref{NC_module}.
\subsubsection{Scaled multiplier $\emph{\textbf{D}}^{(t+1)}$ updating step}
\begin{equation}
	{{\textbf{D}}^{\left( t+1 \right)}}={{\textbf{D}}^{\left( t \right)}}+\textbf{G}{{\textbf{X}}^{\left( t+1 \right)}}-{{\textbf{Z}}^{\left( t+1 \right)}}
	\label{D_solution}
\end{equation}

\subsection{Construction of the proposed SPHR-SAR-Net}
In Section \ref{iterative_solution}, we provided the iterative solution for the high-resolution SAR imaging model based on ADMM. However, the iterative process is slow, memory-intensive, and may not generalize well to new data. Furthermore, manual parameter setting can lead to improper settings, which can negatively affect the imaging results. In this section, we will unroll the iterative solution process into a deep network and train the parameters within SPHR-SAR-Net using a data-driven approach.

\begin{figure*}[!t]
	\centering
	\includegraphics[width=6.5in]{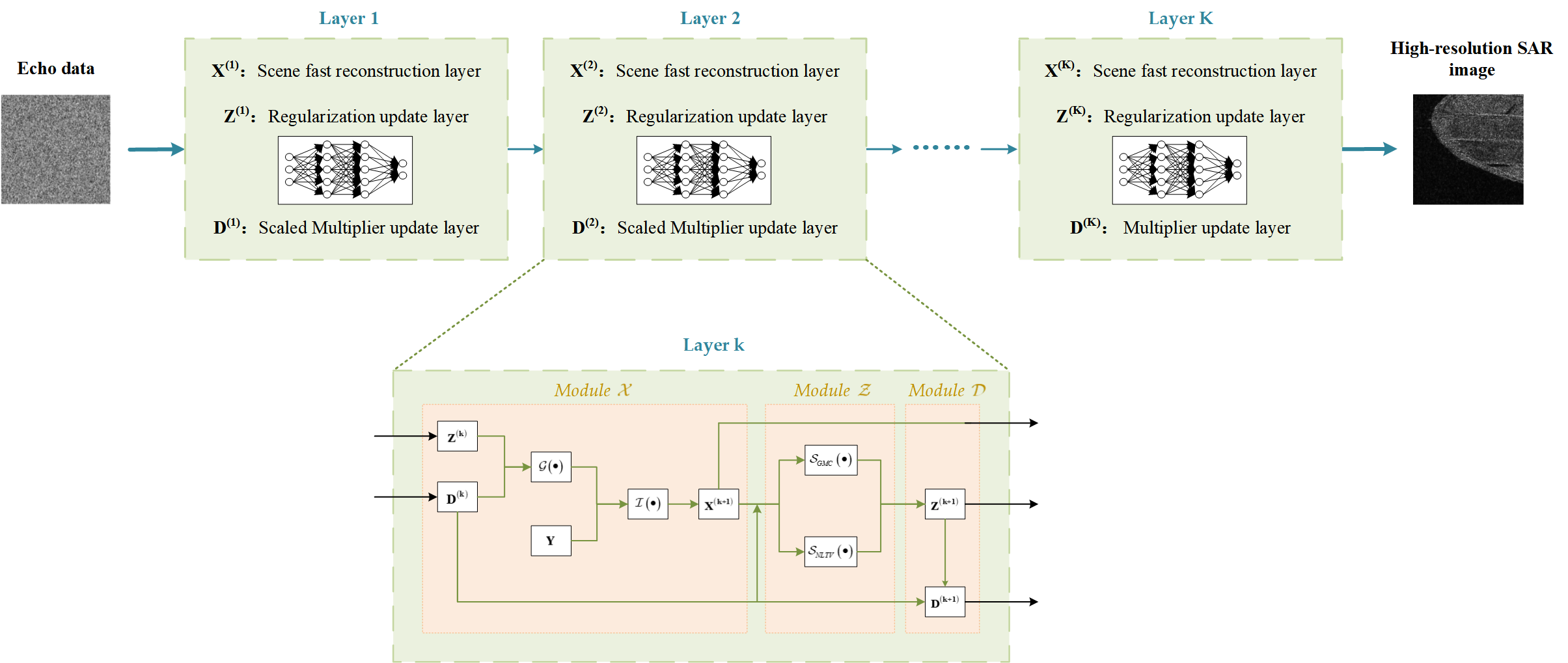}
	\centering
	\caption{Architecture of the proposed SPHR-SAR-Net.}
	\label{fig4}
\end{figure*}

The architecture of the proposed SPHR-SAR-Net is illustrated in Fig. \ref{fig4}. As discussed in Section \ref{iterative_solution}, each updating step in the ADMM iterative solution can be expressed in the form of a closed-form solution, allowing it to be unfolded into an imaging network. As depicted in Fig. \ref{fig4}, the input of the SPHR-SAR-Net is the two-dimensional raw echo data, while the output is the high-resolution SAR images. The SPHR-SAR-Net consists of $K$ layers, with each layer corresponding to an ADMM iterative process. Each layer of the SPHR-SAR-Net contains four modules, which are described in detail below:

\subsubsection{Reconstruction part ${\mathbcal{X}}^{(k+1)}$ module} 
The analytical form of this module corresponds to Equ. (\ref{X_solution}). The inputs of this module are the auxiliary variable ${\textbf{Z}}^{\left( k \right) }$, the scaled multiplier ${\textbf{D}}^{\left( k \right) }$, and the two-dimensional raw echo data $\textbf{Y}$. The output of this module is the reconstructed scene ${\textbf{X}}^{\left( k+1 \right) }$. The penalty parameter ${\rho}^{\left( k+1 \right) }$ is a learnable parameter that is adaptive in different layers. Considering the convergence of the proposed network, the value of the learnable parameter ${\rho}^{\left( k+1 \right) }$ should be between 0 and 1.

\subsubsection{NLTV part ${\mathbcal{Z}_{1}}^{(k+1)}$ module} 
The analytical form of this module corresponds to Equ. (\ref{NLTV_solution}). The inputs of this module are the reconstructed scene ${\textbf{X}}^{\left( k+1 \right) }$ and the scaled multiplier ${\textbf{D}}^{\left( k \right) }$. The output of this module is the auxiliary variable ${\textbf{Z}_{1}}^{\left( k+1 \right) }$ of NLTV. The nonlocal gradient and divergence operators are nonparametric matrix computations. The regularization parameter ${{\lambda }_{NLTV}}^{\left( k+1 \right) }$ and the nonlocal coefficient ${\tau}^{\left( k+1 \right) }$ are learnable parameters that are adaptive in different layers. Here, it is necessary to ensure that ${\tau}^{\left( k+1 \right) }$ is within the theoretical range in Section \ref{NLTV_iteration} so that the network converges.

\subsubsection{NC part ${\mathbcal{Z}_{2}}^{(k+1)}$ module} \label{NC_module}
The analytical form of this module corresponds to Equ. (\ref{NC_solution}). Here, we can apply the ReLU function to represent this non-linear operation. The update of the NC part in the SPHR-SAR-Net can be achieved by:
\begin{multline}
		{\textbf{Z}_{2}}^{\left( k+1 \right)} = \rm{sign} \left( {\textbf{B}^{\left( k+1 \right)}} \right) \odot \\
		 \frac{\text{ReLU}\left( \left| {\textbf{B}^{\left( k+1 \right)}} \right|-\delta  \right)\left[ \vartheta \delta +\text{ReLU}\left( \left| {\textbf{B}^{\left( k+1 \right)}} \right|-\vartheta \delta  \right) \right]}{\left( \vartheta -1 \right)\delta +\text{ReLU}\left( \left| {\textbf{B}^{\left( k+1 \right)}} \right|-\vartheta \delta  \right)}
\end{multline}
where ${\textbf{B}^{\left( k+1 \right)}}={\textbf{X}^{\left( k+1 \right)}}-{\textbf{D}_{2}}^{\left( k \right)}$, $\delta^{\left( k+1\right) }$ is the non-convex threshold, $\vartheta^{\left( k+1\right) }>1$ is the threshold coefficient. $ \left\lbrace \delta^{\left( k+1\right) }, \vartheta^{\left( k+1\right) } \right\rbrace  $ is a set of learnable layer-varying parameters. The inputs of this module are the reconstruction scene ${\textbf{X}}^{\left( k+1 \right) }$ and the scaled multiplier ${\textbf{D}}^{\left( k \right) }$. The output of this module is the auxiliary variable ${\textbf{Z}_{2}}^{\left( k+1 \right) }$ of NC.

\subsubsection{Scaled multiplier part ${\mathbcal{D}}^{(k+1)}$ module} 
The analytical form of this module corresponds to Equ. (\ref{D_solution}). The inputs of this module are the reconstruction scene ${\textbf{X}}^{\left( k+1 \right) }$, the auxiliary variable ${\textbf{Z}}^{\left( k+1 \right) }$ and the scaled multiplier ${\textbf{D}}^{\left( k \right) }$. The output of this module is the scaled multiplier ${\textbf{D}}^{\left( k+1 \right) }$. It can be seen that there are no iteration parameters in this module, so this module is a fixed module and has no learnable parameters in the proposed network.

\subsubsection{Loss function} 
For the training and test tasks, we create a dataset as $\left\lbrace \left( \textbf{Y}_{i}, \textbf{X}_{l,i} \right) \right\rbrace_{i=1}^{{N}_{s}}$, where $\textbf{Y}_{i}$ is the $i \ th$ high-resolution SAR echo data and $\textbf{X}_{l,i}$ is the corresponding label data, ${N}_{s}$ is the size of the dataset. In the above dataset, we select ${N}_{tr}$ data pairs randomly for training and the rest for testing. In the proposed SPHR-SAR-Net, the learnable parameters set is $\left\lbrace {\rho}^{\left( k \right) }, {{\lambda }_{NLTV}}^{\left( k \right) }, {\tau}^{\left( k \right) }, \delta^{\left( k\right) }, \vartheta^{\left( k\right) } \right\rbrace $, which is obtained by minimizing the loss function. The loss function is defined as:
\begin{equation}
	\emph{Loss} = \frac{1}{{N}_{tr}{N}_{p}{N}_{q}} \sum\limits_{i=1}^{{{N}_{tr}}}{\frac{\left\| \left| {{\hat {\textbf{X}}}_{i}} \right|-{\textbf{X}_{l,i}} \right\|_{F}^{2}}{\left\| {\textbf{X}_{l,i}} \right\|_{F}^{2}}}
\end{equation}
where ${\hat {\textbf{X}}}_{i}$ is the output of the proposed SPHR-SAR-Net with the input $\textbf{Y}_{i}$, and $\left\| \cdot  \right\|_{F}^{2}$ is the Frobenius norm.

\section{Experiments and Analysis} \label{experiment}
\subsection{Experimental background}
\subsubsection{Experimental environment}
The implementation of SPHR-SAR-Net is based on PyTorch 1.13.0, and the configuration of the experimental platform consists of an Intel Core i9-12900HX and an NVIDIA RTX 3080 Ti. The optimizer chosen for this task is the Adam optimizer, with the batch size, epoch number, and learning rate set to 4, 100, and 0.0001, respectively. And the layer number is 10.
\subsubsection{Datasets}
For the simulated scenarios, the training and testing datasets consist of random combinations of typical distributed targets with various sizes. The dataset comprises 500 simulated scenes with a size of 512$\times$512, and the echo data are obtained using the inverse imaging operator. Out of these scenes, 450 are randomly selected for training while the remaining 50 are used for testing.

For the real scenarios, both training and testing datasets are obtained by cropping SAR images with a 0.2m azimuth resolution from TerraSAR-X satellite in staring spotlight mode that obtained from Suzhou Industrial Park, China. A total of 500 SAR complex images, each with a size of 512$\times$512, are used to simulate echo data and serve as labels. Out of these, 450 are randomly selected for training, while the remaining 50 are used for testing. Similar to the simulated scenarios, the echo data is also obtained through the inverse imaging operator.

\subsubsection{Evaluation index}
To quantify the advantages of the proposed SPHR-SAR-Net, we conduct a quantitative analysis using various indices. We employ the Equivalent Number of Looks (ENL), Radiometric Resolution ($\gamma$) \cite{xu2021sparse}, and Edge Saving Index (ESI) \cite{chen2019sar} to analyze the reconstructed target results. Additionally, the Peak Signal-to-Noise Ratio (PSNR) and Structural Similarity Index (SSIM) \cite{li2022target} are used to analyze the SAR image reconstruction results under downsampling conditions.

The equivalent number of looks (ENL) and the radiometric resolution are used to evaluate the effectiveness of speckle removal in uniform regions, which is defined by \cite{xu2021sparse}: 
\begin{equation}
	\textbf{ENL}={{\left[ 0.5227  \times \frac{{{\mu }_{{{\hat{\textbf{{X}}}}_{reg}}}}}{{{\sigma }_{{{\hat{\textbf{{X}}}}_{reg}}}}} \right]}^{2}}
\end{equation}
${\mu }_{{{\hat{\textbf{{X}}}}_{reg}}}$ and ${\sigma }_{{{\hat{\textbf{{X}}}}_{reg}}}$  denote the mean and the standard deviation of the reconstructed uniform regions ${{{\hat{\textbf{{X}}}}_{reg}}}$. The larger the values of ENL, the smoother the values of the pixel points in the regions.
  
Furthermore, we can get the definition of radiometric resolution \cite{xu2021sparse}: \begin{equation}
	\bm{\gamma} \left( dB \right)=10\cdot {{\log }_{10}}\left( \frac{{ 0.5227  \cdot {\mu }_{{{\hat{\textbf{{X}}}}_{reg}}}}+{{\sigma }_{{{\hat{\textbf{{X}}}}_{reg}}}}}{{ 0.5227  \cdot {\mu }_{{{\hat{\textbf{{X}}}}_{reg}}}}} \right)
\end{equation}

The edge saving index (ESI) is indicated the edge preserving ability of the SAR imaging method. It is defined as \cite{chen2019sar}:
\begin{equation}
	\textbf{ESI}=\frac{\sum\limits_{i=1}^{N-1}{\sum\limits_{j=1}^{M-1}{\sqrt{{{\left( {{\hat{\textbf{{X}}}}_{i,j}}-{{\hat{\textbf{{X}}}}_{i+1,j}} \right)}^{2}}+{{\left( {{\hat{\textbf{{X}}}}_{i,j}}-{{\hat{\textbf{{X}}}}_{i,j+1}} \right)}^{2}}}}}}{\sum\limits_{i=1}^{N-1}{\sum\limits_{j=1}^{M-1}{\sqrt{{{\left( {{\textbf{X}}_{l}}_{_{i,j}}-{{\textbf{X}}_{l}}_{_{i+1,j}} \right)}^{2}}+{{\left( {{\textbf{X}}_{l}}_{_{i,j}}-{{\textbf{X}}_{l}}_{_{i,j+1}} \right)}^{2}}}}}}
\end{equation}
Higher ESI value means a greater capability for maintaining sharpness of the edges.

The peak signal to noise ratio (PSNR) is a measure of the quality for a reconstructed image relative to the label image, which is described as \cite{li2022target}:
\begin{equation}
	\textbf{PSNR}\left( dB \right)=10\cdot {{\log }_{10}}\frac{\max \left( {{\hat{\textbf{{X}}}}^{2}} \right)}{\left\| \hat{\textbf{X}}-{{\textbf{X}}_{l}} \right\|_{F}^{2}}
\end{equation}
A higher PSNR value indicates that the reconstructed image is of higher quality and closer to the label image.

The structural similarity index (SSIM) is based on the perceived structural similarity of two image and reflects the level of structural similarity between the image, which is defined as \cite{li2022target}:
\begin{equation}
	\textbf{SSIM}=\frac{\left( 2{{\mu }_{\hat{\textbf{X}}}}{{\mu }_{{{\textbf{X}}_{l}}}}+{{C}_{1}} \right)\left( 2{{\sigma }_{\hat{\textbf{X}}{{\textbf{X}}_{l}}}}+{{C}_{2}} \right)}{\left( {{\mu }_{\hat{\textbf{X}}}}^{2}+{{\mu }_{{{\textbf{X}}_{l}}}}^{2}+{{C}_{1}} \right)\left( {{\sigma }_{\hat{\textbf{X}}}}^{2}+{{\sigma }_{{{\textbf{X}}_{l}}}}^{2}+{{C}_{2}} \right)}
\end{equation}
where ${\sigma }_{\hat{\textbf{X}}{{\textbf{X}}_{l}}}$ is the cross covariance of the reconstructed image $\hat{\textbf{X}}$ and the label image ${\textbf{X}}_{l}$. ${C}_{1}$ and ${C}_{2}$ are the constants related to the dynamic range $L$ of the pixel value. In this paper, $L=1$, ${{C}_{1}}={{\left( 0.01\times L \right)}^{2}}$ and ${{C}_{2}}={{\left( 0.03\times L \right)}^{2}}$. The value of SSIM is between 0 and 1, moreover the value closed to 1 indicates that two images are perfectly similar.

\subsection{Experimental procedure and results}
To verify the superiority of the SPHR-SAR-Net in high-resolution SAR imaging, we conduct experiments in both simulated and real SAR scenarios. By adding white Gaussian noise and downsampling to the echo data, it is demonstrated that the proposed method has a strong reconstruction ability even when the echo data is incomplete.

SAR images exhibit coherent speckle, which obscures fine details and makes it challenging to distinguish between objects with similar radar reflectivity. In high-resolution SAR images, the speckle becomes more pronounced, making it more difficult to interpret the information contained in the image. In this paper, the proposed method is compared with the chirp scaling algorithm (CSA) \cite{raney1994precision}, the $\ell_1$-ADMM and NC-NLTV-ADMM \cite{xu2022nonconvex}. For the above methods, the parameter settings are consistent with the empirical values in the article.

\subsubsection{Simulated scenarios}
In this section, we conduct two simulation experiments to validate the effectiveness of our proposed method by comparing it with various approaches for SAR image reconstruction. The simulated scenes are designed as a random combination of distributed targets with varying sizes and positions, maintaining sparsity throughout.

\begin{figure*}[!t]
	\centering
	\subfloat[]{\includegraphics[width=0.3\textwidth]
		{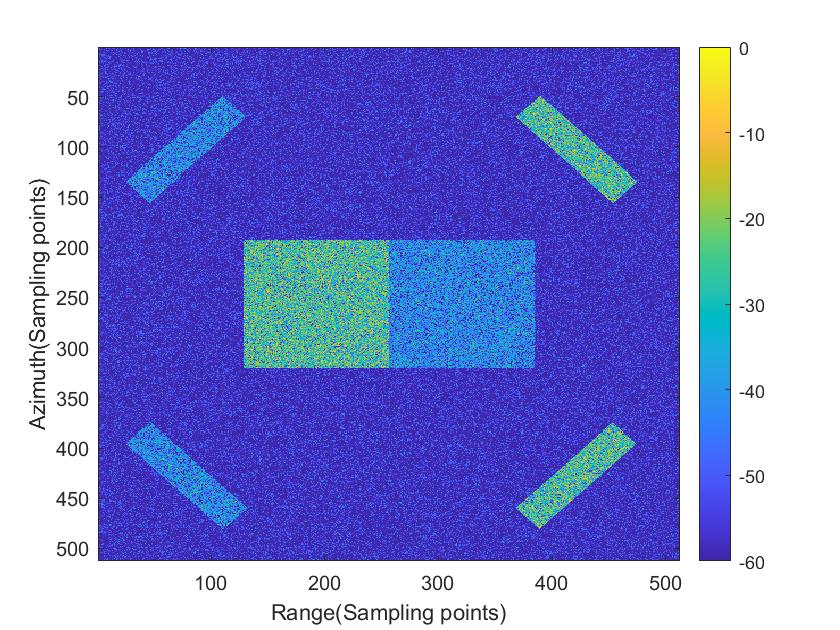}}
	\hfil
	\subfloat[]{\includegraphics[width=0.3\textwidth]
		{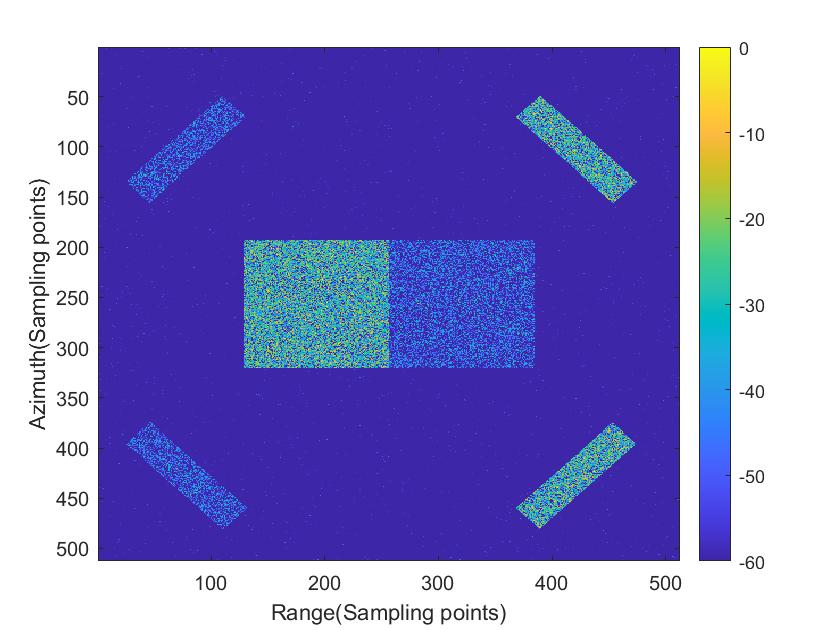}}
	\hfil
	\subfloat[]{\includegraphics[width=0.3\textwidth]
		{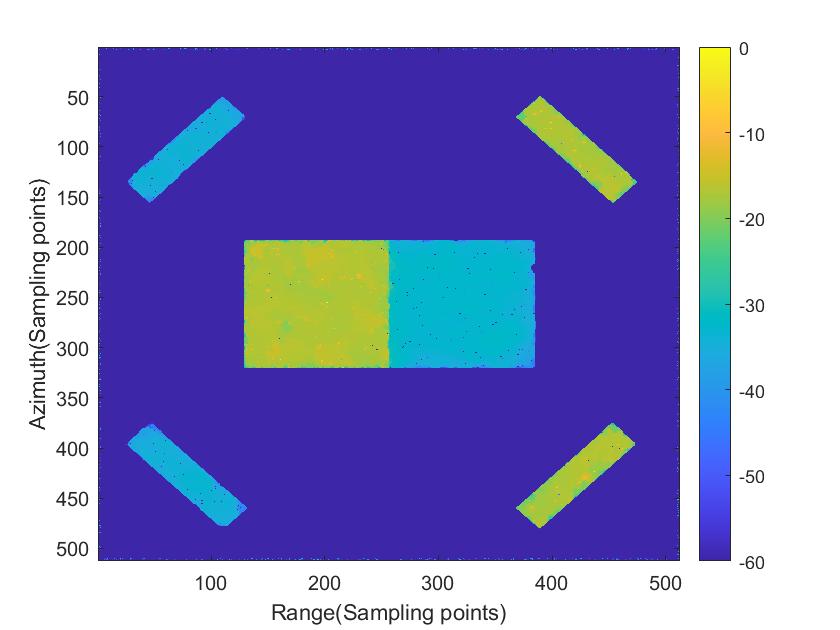}}
	\hfil
	\subfloat[]{\includegraphics[width=0.3\textwidth]
		{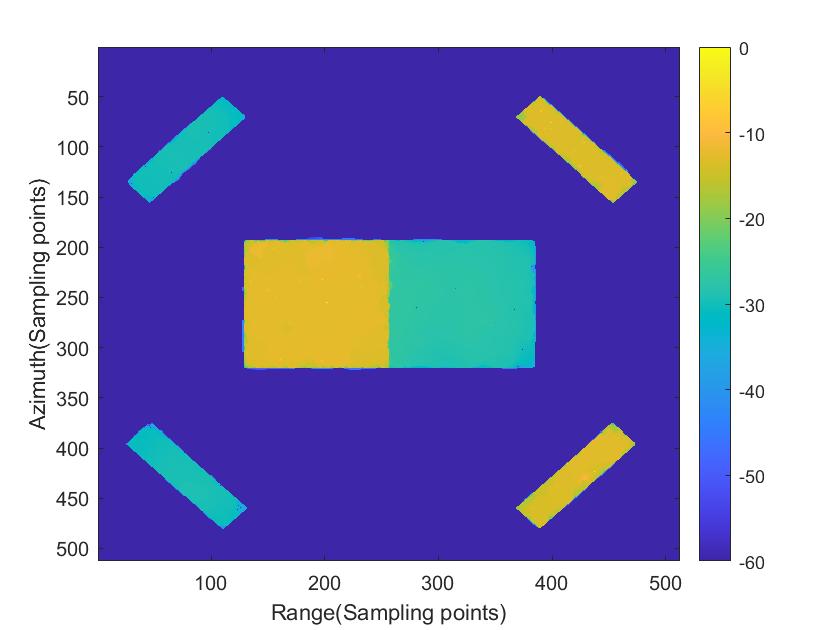}}
	\hfil
	\subfloat[]{\includegraphics[width=0.3\textwidth]
		{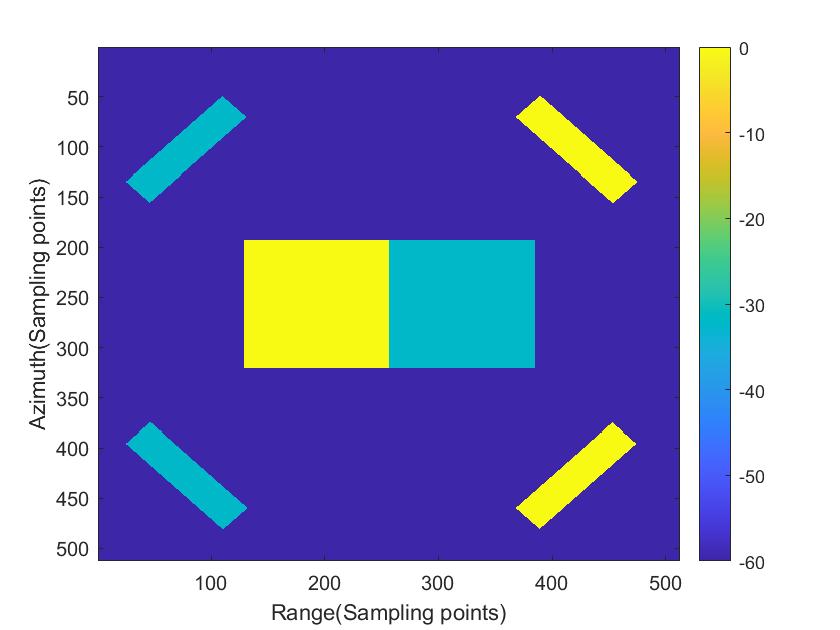}}
	\caption{Simulated scenarios results in different methods where SNR = 5dB. (a) CSA; (b) $\ell_1$-ADMM; (c) NC-NLTV-ADMM; (d) SPHR-SAR-Net; (e) Label.}
	\label{fig5}
\end{figure*}

In the first experiment, we introduced Gaussian white noise to the simulated scene with a signal-to-noise ratio (SNR) of 5dB, and employed various methods to reconstruct the echo data. The reconstruction results are illustrated in Fig. \ref{fig5}. The ENL and $\gamma$ values for CSA and $\ell_1$-ADMM indicate that these methods cannot effectively suppress speckle in SAR images. In comparison, ENL values of the proposed SPHR-SAR-Net is 96.19, and $\gamma$ values is 0.43dB. The traditional iterative method, NC-NLTV-ADMM, has an ENL of 87.46 and a $\gamma$ of 0.44dB, suggesting that its performance is influenced by manually set parameters.

The proposed method exhibits higher PSNR and SSIM values compared to other methods, with PSNR and SSIM values of 16.69dB and 0.8322, respectively. This indicates that our proposed method can accurately reconstruct the scene. Additionally, ADMM iterations have a high time complexity. The ADMM iterative method constrained by $\ell_1$ regularization takes 5.34s, while the method constrained by NC-NLTV-ADMM takes 84.76s. In contrast, the imaging time of the proposed method after training the model is only 1.42s, significantly improving the efficiency of high-resolution SAR imaging.

\begin{figure}[!t]
	\centering
	\subfloat[]{\includegraphics[width=0.2\textwidth]
		{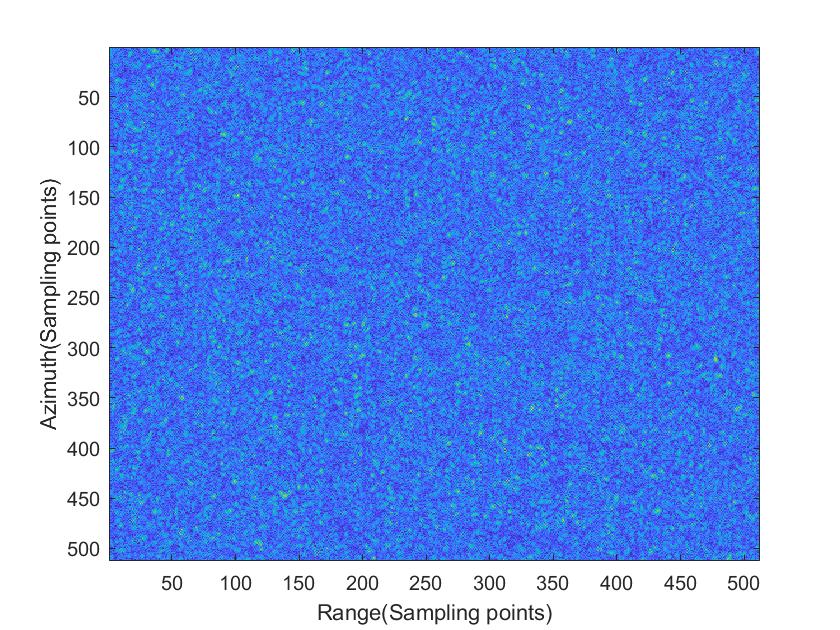}}
	\hfil
	\subfloat[]{\includegraphics[width=0.2\textwidth]
		{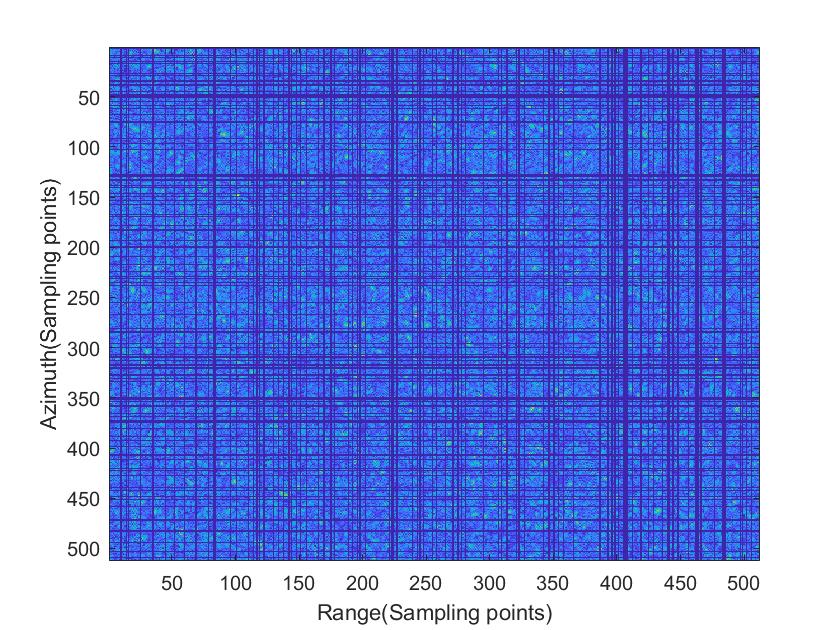}}
	\caption{The echo data after downsampling. (a) Original. (b) After range and azimuth downsampling.}
	\label{fig6}
\end{figure}

\begin{figure*}[!t]
	\centering
	\subfloat[]{\includegraphics[width=0.3\textwidth]
		{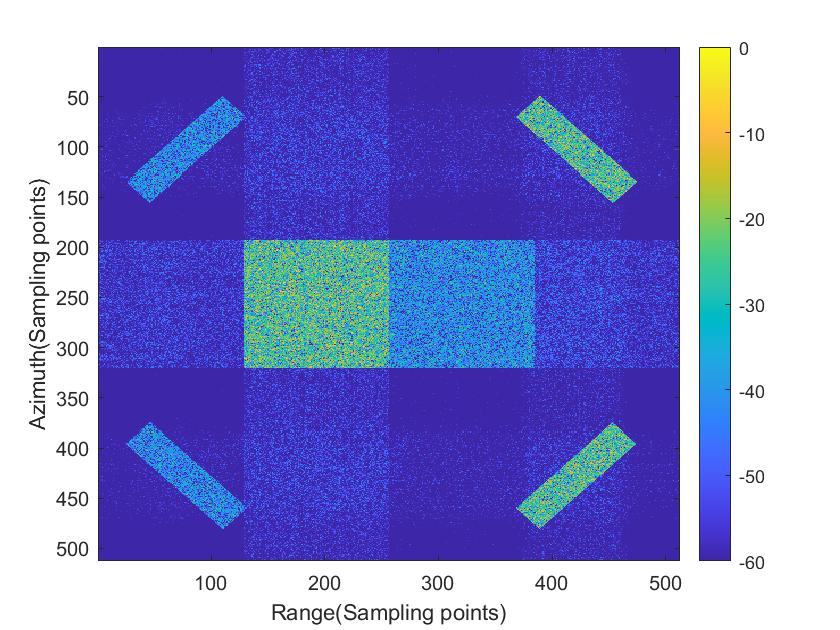}}
	\hfil
	\subfloat[]{\includegraphics[width=0.3\textwidth]
		{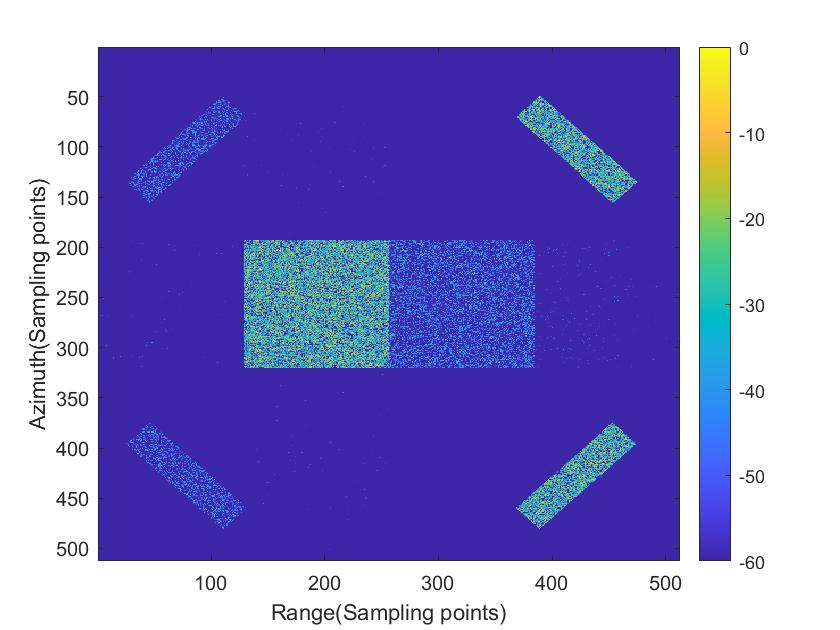}}
	\hfil
	\subfloat[]{\includegraphics[width=0.3\textwidth]
		{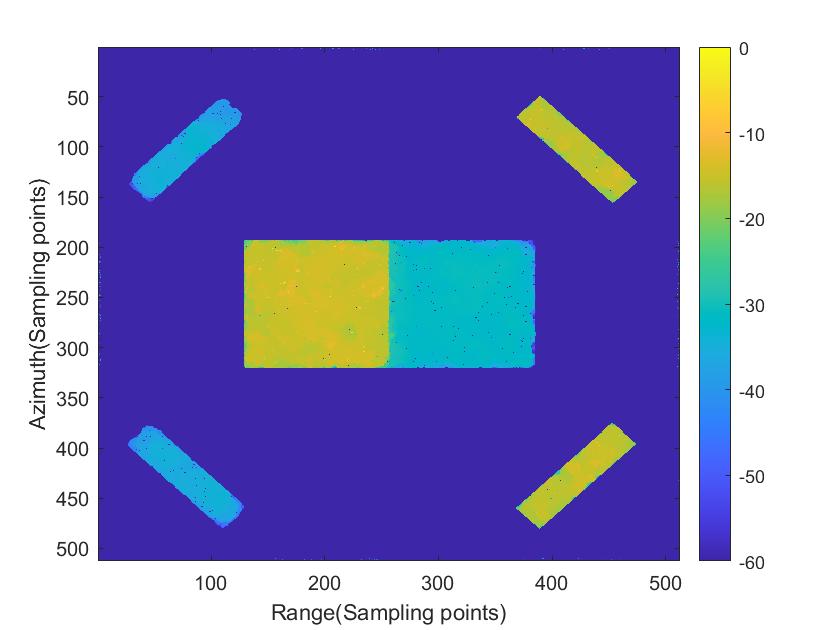}}
	\hfil
	\subfloat[]{\includegraphics[width=0.3\textwidth]
		{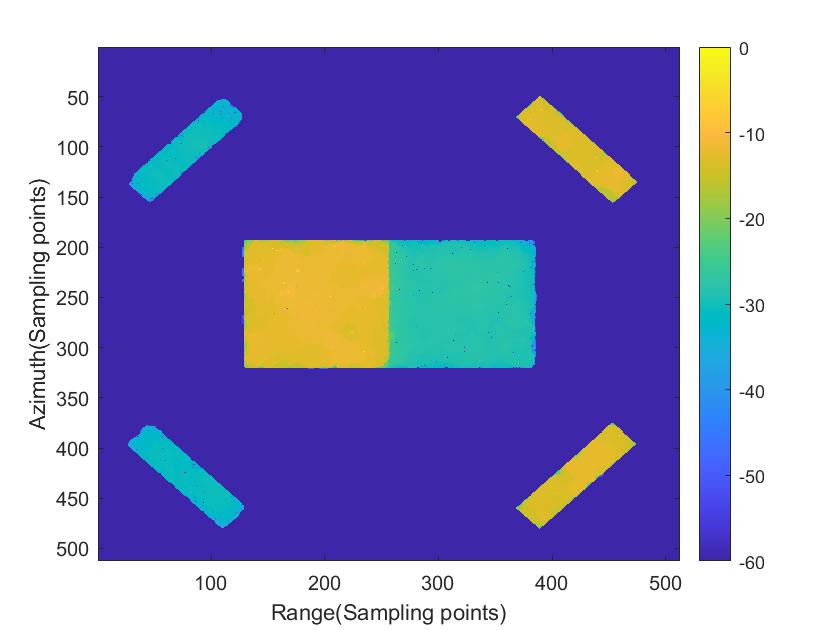}}
	\hfil
	\subfloat[]{\includegraphics[width=0.3\textwidth]
		{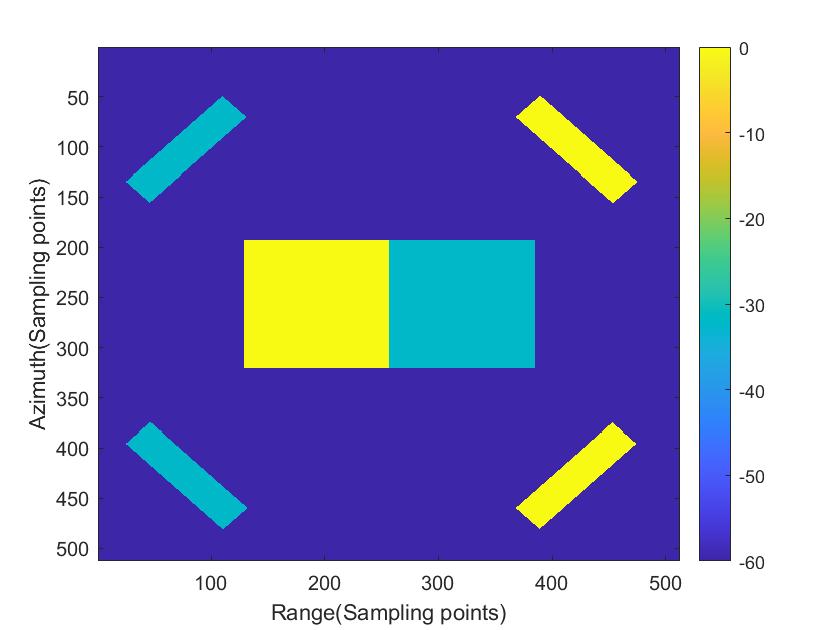}}
	\caption{Simulated scenarios results in different methods where DSR = 81\%. (a) CSA; (b) $\ell_1$-ADMM; (c) NC-NLTV-ADMM; (d) SPHR-SAR-Net; (e) Label.}
	\label{fig7}
\end{figure*}

In the second experiment, we separately downsampled the echo data in the range and azimuth directions. This approach is feasible for reducing data storage and transmission demands in the SAR system. We reduced the data by 10\% in both range and azimuth directions, resulting in a downsampling rate (DSR) of 81\% for the echo data, as illustrated in Fig. \ref{fig6}. We then reconstructed the downsampled echo data using different methods, with the results shown in Fig. \ref{fig7}. Due to the downsampling in range and azimuth, the imaging results display noticeable artifacts and ambiguities, as seen in Fig. \ref{fig7} (a), which can lead to misinterpretations of image outcomes.

In Fig. \ref{fig7} (b), $\ell_1$-ADMM employs sparse signal processing technology to effectively eliminate the ambiguities arising from downsampling. However, it does not remove the speckle, resulting in ENL and $\gamma$ values of 0.41 and 4.07dB, respectively. Moreover, this method preserves fine details and edges while reducing the staircase effect, ensuring the accuracy of high-resolution SAR imaging. Owing to the influence of manually set iteration parameters, the SAR imaging results of NC-NLTV-ADMM are inferior to those of the proposed method in terms of speckle filtering, edge preservation, and image reconstruction accuracy. Additionally, the iteration time of this method is significantly longer than that of the proposed method.

The simulation results demonstrate that the proposed method enhances imaging accuracy in high-resolution SAR imaging and effectively removes additive white Gaussian noise and speckle noise in high-resolution SAR images. Furthermore, the deep unrolling approach of the SPHR-SAR-Net significantly reduces the imaging time.

\begin{table*}[!t]  
	\centering  
	\renewcommand{\arraystretch}{1.5} %控制行高  
	\setlength{\tabcolsep}{5mm}
	\fontsize{8}{9}\selectfont  
	\begin{threeparttable}  
		\caption{index analysis of different methods for simulation experiment scenarios}    
		\begin{tabular}{cccccccc}  
			\toprule  
			{\bf Name}& {\bf Method} & {\bf{ENL}} & {$\bm{\gamma}\left( dB \right) $} &	{\bf{ESI}} & {\bf{PSNR}$\left(dB\right)$} & {\bf{SSIM}} & {\bf{Time(s)}}    \cr 
			\midrule
			\multirow{5}{*}{\bf{SNR = 5dB}}  &
			CSA & 0.99 & 3.01 &	/ &	12.75 & 0.0521 &	0.16     \cr 
			~ & $\ell_1$-ADMM & 0.54 & 3.70 & / & 12.24 &	0.6597 & 5.34 \cr
			~ & NC-NLTV-ADMM & 87.46 & 0.44 & 0.4096 &	15.22 & 0.8640 & 84.76 \cr
			~ & SPHR-SAR-Net & {\bf 96.19 }& {\bf 0.43} & {\bf 0.4439} & {\bf 16.69} & {\bf 0.8322} & {\bf 1.42}  \cr
			\midrule
			\multirow{5}{*}{\bf{DSR = 81\%}}  &
			CSA & 1.07 & 2.93 &	/ &	12.91 & 0.1046 & 0.15     \cr 
			~ & $\ell_1$-ADMM & 0.41 & 4.07 & / & 12.21 &	0.7963 & 5.44 \cr
			~ & NC-NLTV-ADMM & 76.56 & 0.47 &	0.3928 &	15.82 & 0.8228 & 82.34 \cr
			~ & SPHR-SAR-Net & {\bf 86.48 }& {\bf 0.44} & {\bf 0.4436} & {\bf 20.24} & {\bf 0.6044} & {\bf 1.62}  \cr
			\bottomrule  
		\end{tabular}  
	\end{threeparttable}  
	\label{Tab. 1}
\end{table*} 

\subsubsection{Real SAR scenarios}
\begin{figure*}[!t]
	\centering
	\subfloat[]{\includegraphics[width=0.35\textwidth]
		{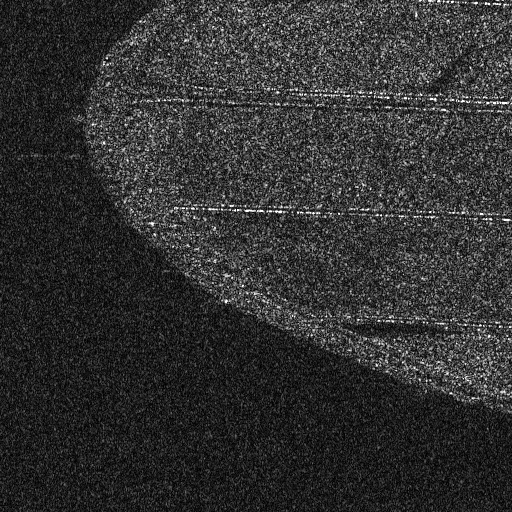}}
	\hfil
	\subfloat[]{\includegraphics[width=0.35\textwidth]
		{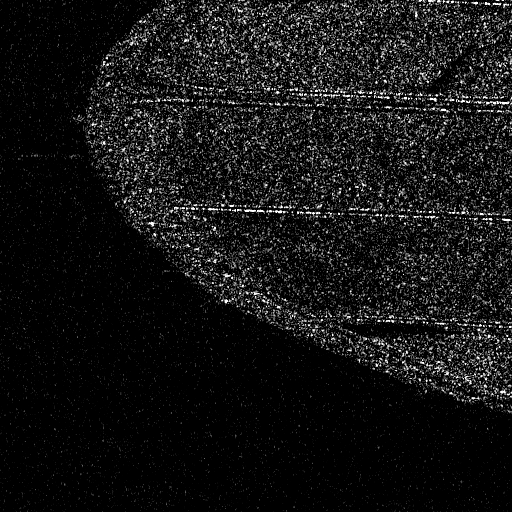}}
	\hfil
	\subfloat[]{\includegraphics[width=0.35\textwidth]
		{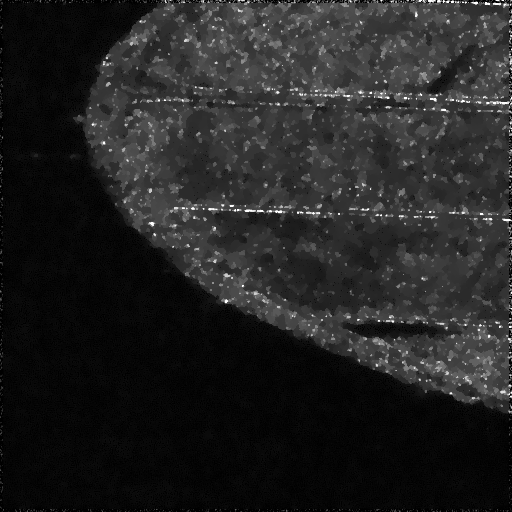}}
	\hfil
	\subfloat[]{\includegraphics[width=0.35\textwidth]
		{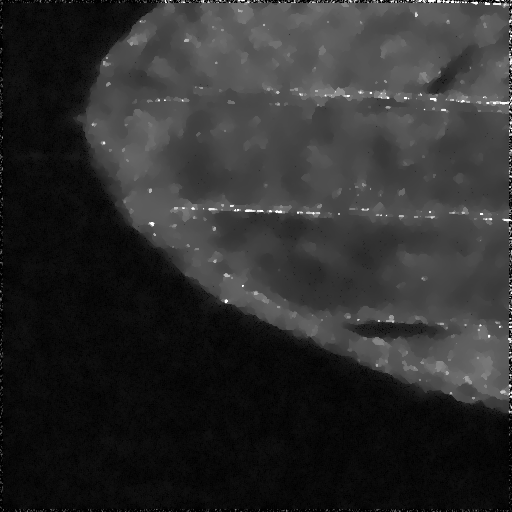}}

	\caption{Real scenarios results of Scene 1. (a) CSA; (b) $\ell_1$-ADMM; (c) NC-NLTV-ADMM; (d) SPHR-SAR-Net.}
	\label{fig8}
\end{figure*}

\begin{figure*}[!t]
	\centering
	\subfloat[]{\includegraphics[width=0.35\textwidth]
		{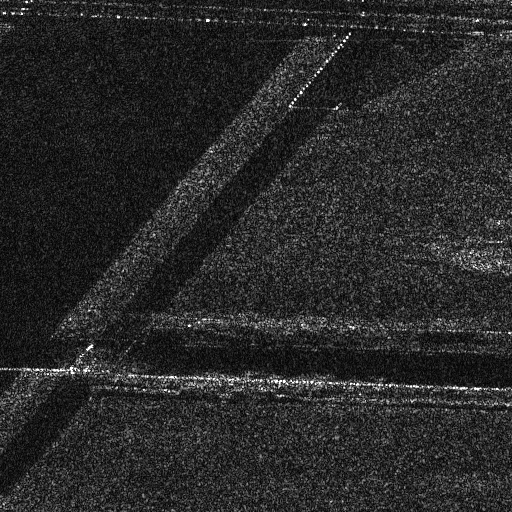}}
	\hfil
	\subfloat[]{\includegraphics[width=0.35\textwidth]
		{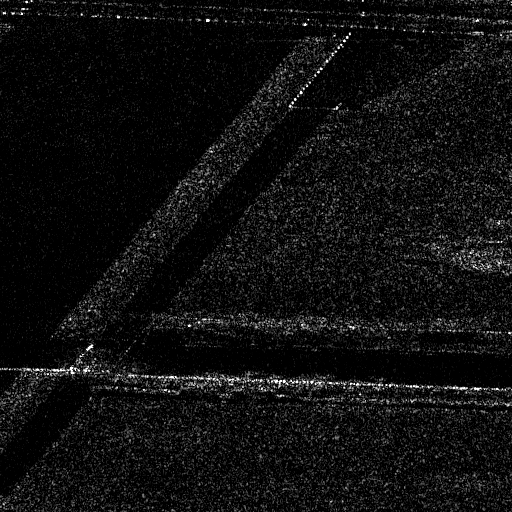}}
	\hfil
	\subfloat[]{\includegraphics[width=0.35\textwidth]
		{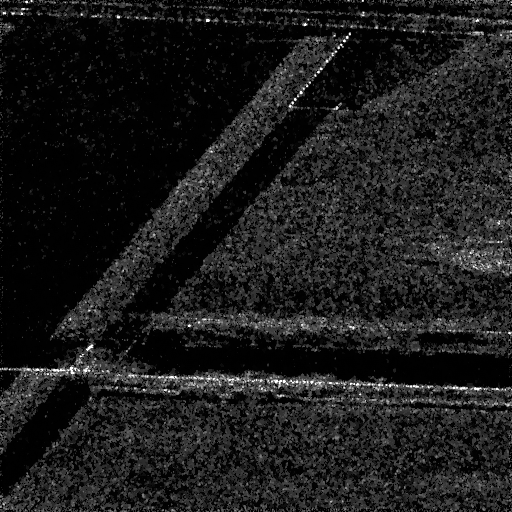}}
	\hfil
	\subfloat[]{\includegraphics[width=0.35\textwidth]
		{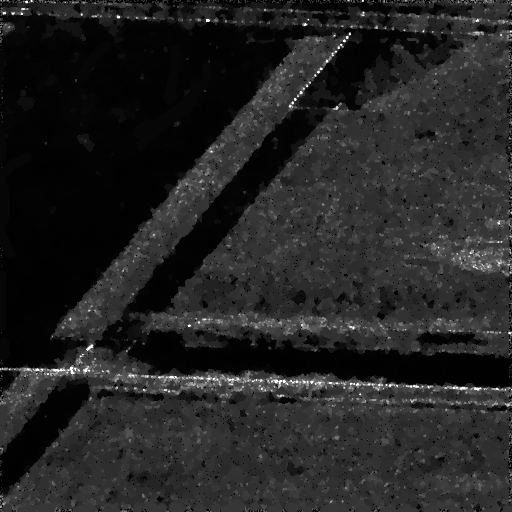}}
	
	\caption{Real scenarios results of Scene 2. (a) CSA; (b) $\ell_1$-ADMM; (c) NC-NLTV-ADMM; (d) SPHR-SAR-Net; (e) Label.}
	\label{fig9}
\end{figure*}

In this subsection, we evaluate the effectiveness of the proposed method using two scenes for SAR image reconstruction by comparing it with different methods. The first scene features a water bank, while the second scene depicts the junction of farmland. The imaging results for the two scenes are presented in Fig. \ref{fig8} and Fig. \ref{fig9}, respectively. We then perform index analysis on the imaging results obtained from different methods, with the results displayed in Tables \ref{Tab. 2} and \ref{Tab. 3}.

Compared with NC-NLTV-ADMM, the proposed method is more effective at removing speckle noise and smoothing distributed targets. The ENL values for SPHR-SAR-Net are 80.07 and 60.59, while those for NC-NLTV-ADMM are 48.56 and 19.00 The $\gamma$ values for SPHR-SAR-Net are 0.46dB and 0.52dB, while those for NC-NLTV-ADMM are 0.58dB and 0.89dB. Compared with $\ell_1$-ADMM and NC-NLTV-ADMM, SPHR-SAR-Net achieves higher PSNR and SSIM values, suggesting that it has a higher similarity with the label images. The PSNR values for SPHR-SAR-Net are 24.63dB and 32.96dB, while the SSIM values are 0.5969 and 0.6921. Furthermore, SPHR-SAR-Net significantly improves imaging speed compared to the ADMM iterative method. The proposed method's time consumption is 60 times less than that of NC-NLTV-ADMM, enabling rapid high-resolution SAR imaging from echo data.

\begin{table*}[!t]  
	\centering  
	\renewcommand{\arraystretch}{1.5} %控制行高  
	\setlength{\tabcolsep}{5mm}
	\fontsize{8}{9}\selectfont  
	\begin{threeparttable}  
		\caption{ENL, $\gamma$, ESI and Time of different methods for real SAR scenarios}    
		\begin{tabular}{ccccccccc}  
			\toprule  
			\multirow{2}{*}{\bf{Method}} &  
			\multicolumn{2}{c}{\bf{ENL}} & \multicolumn{2}{c}{$\bm{\gamma}\left( dB \right) $} & \multicolumn{2}{c}{\bf{ESI}}&
			\multicolumn{2}{c}{\bf{Time(s)}}\cr  
			\cmidrule(lr){2-3} \cmidrule(lr){4-5}  \cmidrule(lr){6-7} \cmidrule(lr){8-9}
			& \bf{Scene 1} & \bf{Scene 2} & \bf{Scene 1} & \bf{Scene 2} & \bf{Scene 1} & \bf{Scene 2}  & \bf{Scene 1} & \bf{Scene 2}  \cr  
			\midrule  
			CSA & 0.97 & 0.92 &	3.03 & 3.10 & / & / & 0.19 & 0.18   \cr 
			$\ell_1$-ADMM & 0.16 & 0.24 & 5.48 & 4.82 &	/ & / & 5.43 & 6.76\cr
			NC-NLTV-ADMM & 48.56 &	19.00 & 0.58 & 0.89 & 0.4136 & 0.3977 & 85.21 & 83.98\cr
			SPHR-SAR-Net & {\bf 80.07 }& {\bf 60.59} & {\bf 0.46} & {\bf 0.52} & {\bf 0.5190} & {\bf 0.3156} & {\bf 1.42} & {\bf 1.22} \cr
			\bottomrule  
		\end{tabular} 
		\label{Tab. 2} 
	\end{threeparttable}  
\end{table*}

\begin{table*}[!t]  
	\centering  
	\renewcommand{\arraystretch}{1.5} %控制行高  
	\setlength{\tabcolsep}{6mm}
	\fontsize{8}{9}\selectfont  
	\begin{threeparttable}  
		\caption{PSNR and SSIM of different methods for real SAR scenarios}    
		\begin{tabular}{ccccc}  
			\toprule  
			\multirow{2}{*}{\bf{Method}} &  
			\multicolumn{2}{c}{\bf{PSNR}$\left(dB\right)$}& 
			\multicolumn{2}{c}{\bf{SSIM}}\cr  
			\cmidrule(lr){2-3} \cmidrule(lr){4-5}  
			& \bf{Scene 1} & \bf{Scene 2} & \bf{Scene 1} & \bf{Scene 2}  \cr  
			\midrule  
			$\ell_1$-ADMM & 20.01 & 26.66 & 0.1437 & 0.3239\cr
			NC-NLTV-ADMM & 23.05 & 30.06 & 0.4742 & 0.4880\cr
			SPHR-SAR-Net & {\bf 24.63 }& {\bf 32.96} & {\bf 0.5969} & {\bf 0.6921} \cr
			\bottomrule  
		\end{tabular} 
		\label{Tab. 3} 
	\end{threeparttable}  
\end{table*}

\section{Conclusions} \label{conclusion}
In this paper, we have proposed a Superpixel High-Resolution SAR Imaging Network, or SPHR-SAR-Net, which employs nonlocal total variation and non-convex compound regularization terms to achieve precise constraints on superpixels in high-resolution SAR images. The architecture of SPHR-SAR-Net is derived from the ADMM solver for this compound regularization, with layer-specific parameters learned in an end-to-end fashion. Furthermore, SPHR-SAR-Net integrates high-resolution SAR imaging modalities into a deep unfolded network.

The effectiveness of the proposed method has been demonstrated through experiments on simulated and real SAR scenarios. When compared to traditional filtering methods, SPHR-SAR-Net performs imaging from echo data, preserving phase information while achieving superior imaging results. In comparison to iterative-based regularization methods, SPHR-SAR-Net significantly improves both imaging quality and speed. As a result, SPHR-SAR-Net can be applied to fast and accurate imaging of high-resolution SAR.

% References
\bibliographystyle{IEEEtran}
\bibliography{reference}

\end{document}